\documentclass[12pt,preprint]{aastex}

\slugcomment{DRAFT: \today}

\shorttitle{The $s$ process in massive stars}
\shortauthors{}

\begin{document}

\title{The $s$ process in massive stars at low metallicity. Effect of primary
$^{14}$N from fast rotating stars.} 
\author{M. Pignatari\altaffilmark{1,2}, 
R. Gallino\altaffilmark{3,4}, 
G. Meynet\altaffilmark{5}, 
R. Hirschi\altaffilmark{1,6},
F. Herwig\altaffilmark{1,7}, 
M. Wiescher\altaffilmark{2},
}
\altaffiltext{1}{Keele University, Keele, Staffordshire ST5 5BG, 
United Kingdom; marco@astro.keele.ac.uk, r.hirschi@epsam.keele.ac.uk}
\altaffiltext{2}{JINA, University of Notre Dame, Notre Dame, IN 46556, USA;
wiescher.1@nd.edu}
\altaffiltext{3}{Dipartimento di Fisica Generale, Universit\'a di Torino, 
Via Pietro Giuria 1, Torino 10125, Italy; gallino@ph.unito.it}
\altaffiltext{4}{Center for Stellar and Planetary Astrophysics,
School of Mathematical Sciences, Monash University, PO Box 28,
Victoria 3800, Australia.}
\altaffiltext{5}{Geneva Observatory, CH-1290 Sauverny, Switzerland; 
georges.meynet@obs.unige.ch}
\altaffiltext{6}{IPMU, University of Tokyo, Kashiwa, Chiba 277-8582, Japan.}
\altaffiltext{7}{Department of Physics \& Astronomy, University of Victoria, Victoria, BC, 
V8P5C2 Canada; fherwig@uvic.ca}

\begin{abstract}
The goal of this paper is to analyze the 
impact of a primary neutron source on the $s-$process nucleosynthesis 
in massive stars at halo metallicity.
Recent stellar models including rotation at very low metallicity predict a
strong production of primary $^{14}$N.
Part of the nitrogen produced in the H$-$burning shell diffuses by rotational
mixing into the He core where it is converted to $^{22}$Ne 
providing additional neutrons for the $s$ process. 
We present nucleosynthesis calculations for a 25 M$_{\odot}$ star at 
[Fe/H]
= $-$3, $-$4, 
where in the convective core He$-$burning about 0.8 \% in mass is made
of primary $^{22}$Ne. 
The usual weak $s-$process shape is changed by the additional 
neutron source with a peak between Sr and Ba, where
the $s-$process yields increase by orders of magnitude with
respect to the yields obtained without rotation.
Iron seeds are fully consumed and the maximum production of Sr, Y and Zr is 
reached. On the other hand, the $s-$process efficiency beyond Sr and the ratio 
Sr/Ba are strongly affected by the amount of $^{22}$Ne and by nuclear 
uncertainties, first of all by the $^{22}$Ne($\alpha$,n)$^{25}$Mg reaction. 
Finally, assuming that $^{22}$Ne is primary in the considered metallicity range,
the $s-$process efficiency decreases with metallicity  
due to the effect of the major neutron poisons $^{25}$Mg and $^{22}$Ne.
This work represents a first step towards the study of primary neutron
source effect in fast rotating massive stars, and its 
implications are discussed in the light of
spectroscopic observations of heavy elements at halo metallicity.
\end{abstract}


\section{Introduction}
\label{sec:intro}

Very metal$-$poor halo stars have formed from matter that was only enriched
by massive stars ejecta \citep[e.g.][hereafter Tr04]{travaglio:04}. 
Their surface compositions thus reflect the nucleosynthesis occurring in the 
first generations of massive stars. 
A puzzling feature observed in a significant number of
stars is the high Sr enrichment correlated with low Ba abundance 
\citep[e.g.][]{truran:02, aoki:05}, with the observed Sr/Ba ratio 
spreading over more than two orders of magnitude.
They are not explained by the standard $r$ process 
\citep[rapid neutron capture process,][]{burbidge:57}.
In particular, the $r-$process is responsible for the production of about 20 \%
of solar Ba and 12 \% of solar Sr, causing a [Sr/Ba]$_{\rm r}$ = $-$0.22 
\citep[][]{sneden:08}.
On the other hand, stars like CS 22949$-$037 ([Fe/H] = $-$4) 
and the most metal$-$poor star known
HE 1327$-$2326 ([Fe/H] = $-$5.45) are 
Sr$-$Y rich, Ba poor and with high [Sr/Ba] 
\citep[for details see][respectively]{depagne:02, aoki:06}.
Another process must be responsible for the Sr observations, which is
related to a similar observed Y and Zr overproduction with respect to Ba 
\citep[e.g., Tr04,][]{montes:07}.
Asymptotic giant branch (AGB) stars do not have time
to contribute to the Galactic chemical evolution (GCE) 
at very low metallicity.
The standard weak $s$ process (slow neutron capture process) in massive stars can neither be the
solution for the Sr$-$Y$-$Zr overproduction. Indeed,
the secondary nature \footnote{The production of secondary (primary) isotopes
does (does not) depend on the initial metallicity of the star.
}
of the main neutron source 
$^{22}$Ne progressively reduces the $s-$process 
efficiency with decreasing metallicity. 
In particular, for [Fe/H] $\le$ $-$2 primary neutron poisons 
capture most of the neutrons causing an even faster decrease
of the $s-$process yields with decreasing metallicity \citep[][]{raiteri:92}.   
Tr04 propose that an unknown primary neutron capture process in massive stars 
is responsible for the production of Sr$-$Y$-$Zr, and provide an evaluation 
of the amount of Sr, Y and Zr that should be produced in order to explain 
observations at very low metallicity: about 10 \% of the solar Sr and 
20 \% of the solar Y and Zr.
They called this process light element primary process (LEPP). 
The unknown component could also be produced by explosive nucleosynthesis 
in massive stars.
If in core collapse supernovae the 
$\alpha-$rich freezout is not complete,
the final explosive nucleosynthesis in high$-$entropy neutrino winds 
from a forming neutron star may be significant for the Sr peak elements
\citep[][]{woosley:92}.
More recently, \cite{farouqi:08} and \cite{qian:07} generally confirm 
the feasibility of this nucleosynthesis scenario in high$-$entropy 
neutrino winds.\\
Recent results from massive stars calculations at low metallicity 
including rotation 
could have important implications in this discussion.
In massive stars at very low metallicity, rotation boosts the formation 
of primary $^{14}$N. 
According to \cite{meynet:06} and \cite{hirschi:07}, due to rotational mixing 
primary $^{12}$C produced in the convective He core is brought in contact with 
the bottom of the convective H shell, where it is converted into 
$^{14}$N via the CN cycle.
Part of the $^{14}$N is mixed in the envelope, and part is mixed 
down or engulfed in the He core \citep[e.g.][]{meynet:06}.
A more recent model by \cite{hirschi:08} for a 20 M$_{\odot}$ star predicts that 
H shell ashes are directly processed by the convective core He$-$burning, 
causing an even higher amount of primary 
$^{14}$N in the He core than in the previous models.
During He$-$burning, nitrogen is rapidly converted 
to $^{22}$Ne by the $\alpha-$capture sequence 
$^{14}$N($\alpha$,$\gamma$)$^{18}$F($\beta^+\nu$)$^{18}$O($\alpha$,$\gamma$)$^{22}$Ne.
Close to the He exhaustion, $^{22}$Ne($\alpha$,n)$^{25}$Mg is activated 
providing neutrons for the $s$ process. 
\cite{hirschi:08} quantified that for a 20 M$_{\odot}$ star 
with [Fe/H] = $-$4.3 the primary $^{22}$Ne in the He core before the 
$s-$process activation is about 0.8 \% in mass,
about 200 times more than in the case with no rotation.  \\
Spectroscopic observations by \cite{spite:05} and \cite{israelian:04}
indicate a primary production of nitrogen over a large metallicity range 
for [Fe/H] $>$ $-$4.1, 
and also at lower metallicities strong N enrichment has been observed in stars
correlated with Sr enrichment \citep[e.g. HE 1327$-$2326,][]{frebel:05}.
\cite{chiappini:06} show that the primary nature of nitrogen can be explained 
at low metallicity by using the yields of fast rotating massive stars. 
Since the production of primary nitrogen is directly related to the 
$^{22}$Ne available in the He core, 
we may also expect that the additional $^{22}$Ne is produced in a primary way
over a large metallicity range. \\
In the present work, our aim is to investigate possible consequences of 
massive star models including rotation
on the synthesis of $s-$process elements.
In \S \ref{sec:disc} we present $s-$process 
nucleosynthesis calculations in the 
He core and in the following convective C$-$burning shell of a 25 M$_{\odot}$ 
at [Fe/H] = $-$3, $-$4. Final discussion and conclusions are given in 
\S \ref{sec:concl}.

\section{Calculations and results}
\label{sec:disc}

\begin{figure}[!h]
%\centering
\resizebox{18pc}{!}{\rotatebox{-90}
{\includegraphics{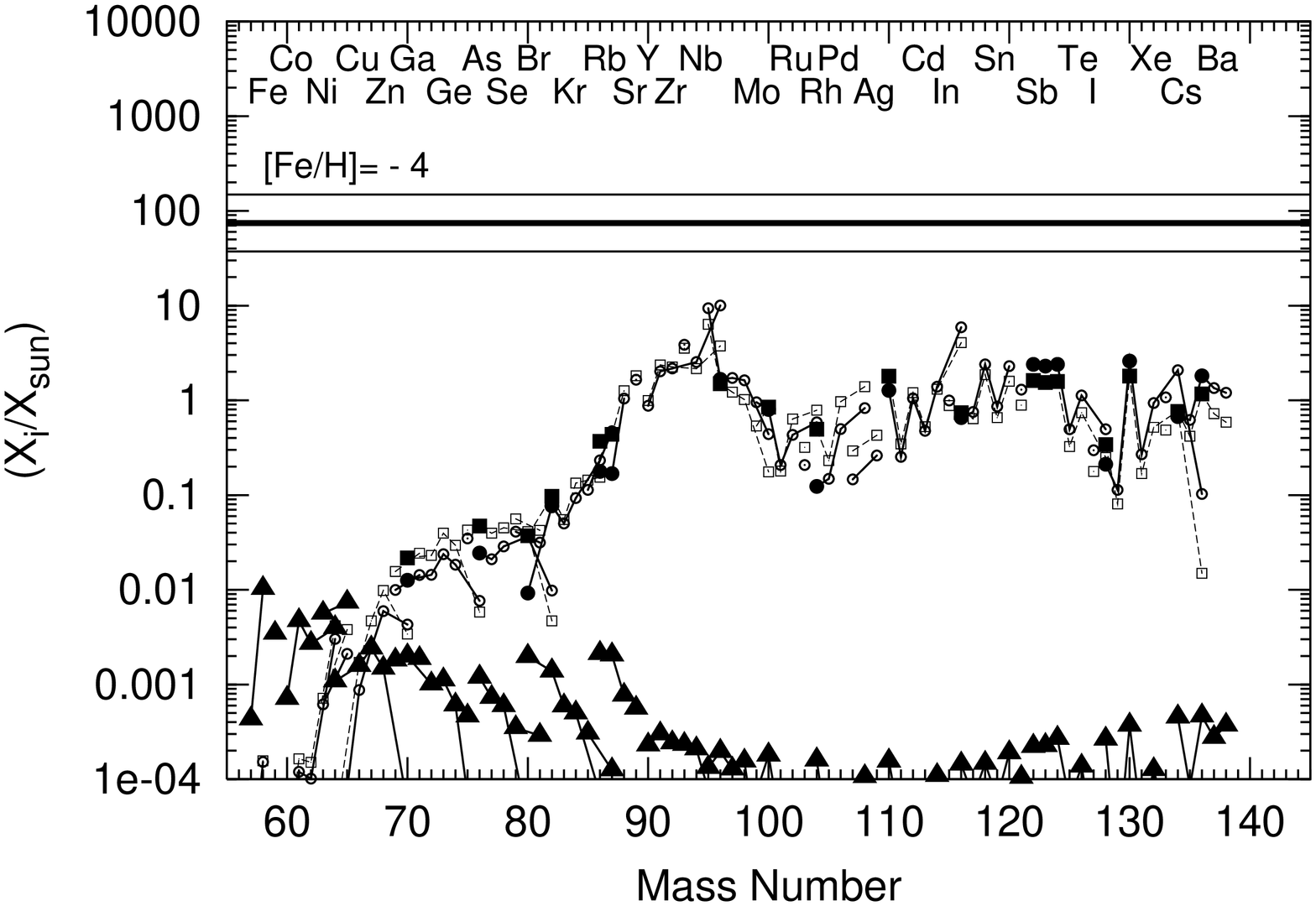}}}
\hfill
\resizebox{18pc}{!}{\rotatebox{-90}
{\includegraphics{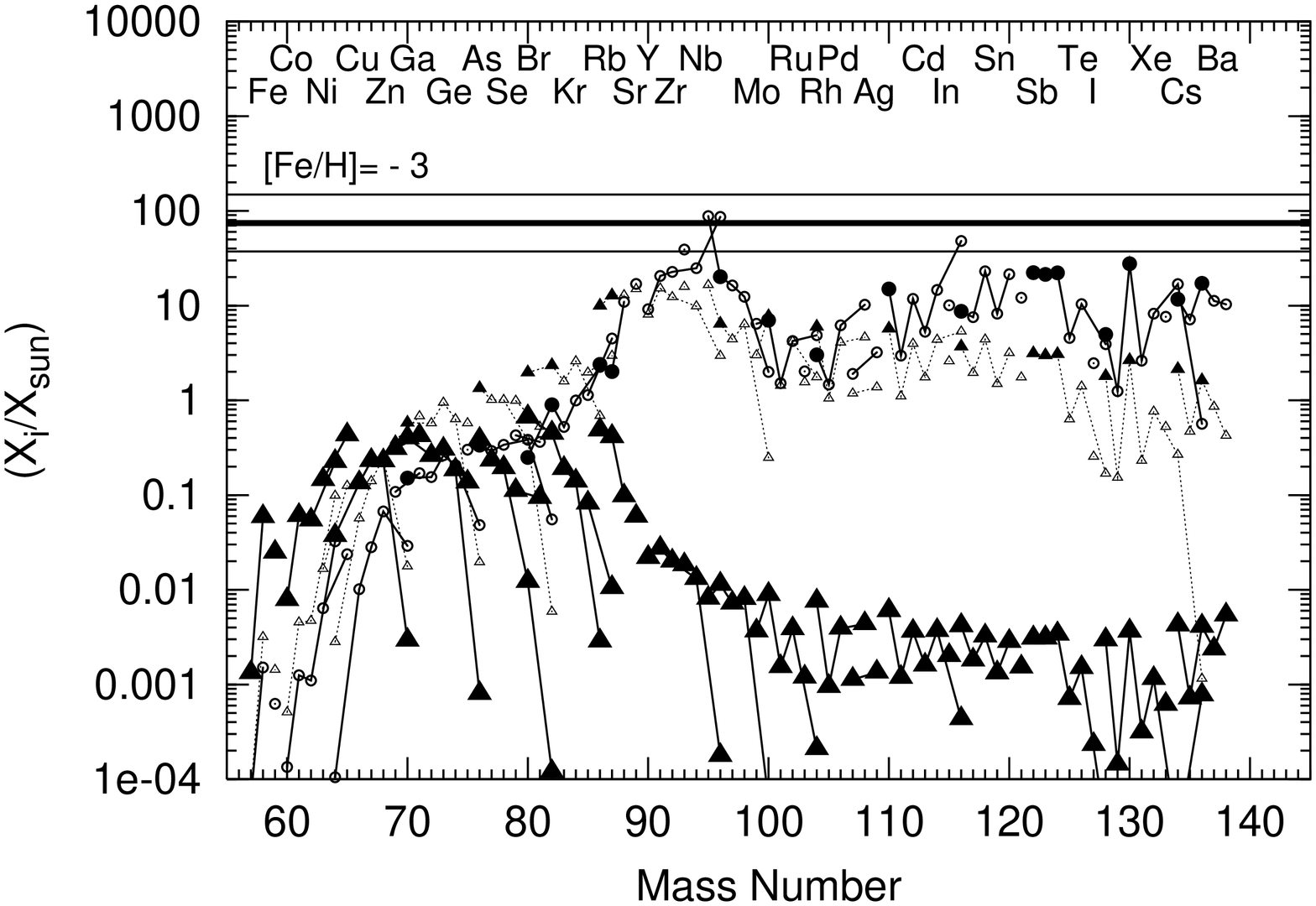}}} 
\caption{$Top Panel$: $s-$process  
distributions between $^{57}$Fe and $^{138}$Ba normalized to solar 
for the 25 M$_{\odot}$ and [Fe/H] = $-$4 at the end of the convective C$-$burning shell.
The horizontal lines correspond to 
the $^{16}$O overabundance in the C shell (thick line), multiplied and 
divided by two (thin lines). 
Isotopes of the same element are
connected by a line.
The cases presented are the following: $i$) standard (black triangles), 
$X$($^{22}$Ne)$_{\rm ini}$ = 5.21~$\times$~10$^{-5}$;
$ii$) $X$($^{22}$Ne)$_{\rm ini}$ = 5.0~$\times$~10$^{-3}$ 
(open squares, full squares for the $s-$only isotopes); 
$iii$) $X$($^{22}$Ne)$_{\rm ini}$ = 1.0~$\times$~10$^{-2}$ 
(open circles, full circles for the $s-$only isotopes).
$Bottom Panel$: the same as Top Panel, at [Fe/H] = $-$3: 
$i$) standard (black triangles), 
$X$($^{22}$Ne)$_{\rm ini}$ = 2.12~$\times$~10$^{-4}$;
$ii$) $X$($^{22}$Ne)$_{\rm ini}$ = 2.0~$\times$~10$^{-3}$ (small triangles); 
$iii$) $X$($^{22}$Ne)$_{\rm ini}$ = 1.0~$\times$~10$^{-2}$ (small circles).}
\label{m25}
\end{figure}

We present $s-$process calculations for a 25 M$_{\odot}$ star at [Fe/H]= $-$3, $-$4. 
A post$-$processing code has been used to follow the $s$ process in the 
convective He core phase and in the convective C shell phase 
\citep[]{raiteri:91b}.
In particular, the C shell burns over the ashes of the previous He core. 
In this phase the main neutron source for neutron capture is the $^{22}$Ne 
left after He$-$burning.
During the supernova explosion, most of the convective C shell region 
is ejected unchanged carrying the $s-$process signature of the 
pre$-$explosive phase.   
For a 25 M$_{\odot}$, indeed, the material up to
3$-$3.5 M$_{\odot}$ is further processed and the s yields are destroyed.
Between 3$-$3.5 M$_{\odot}$ and 6$-$6.5 M$_{\odot}$ the O rich and 
$s-$process rich material is ejected almost unchanged by the explosion 
\citep[e.g.][]{woosley:02}.\\
The nuclear network has been updated, as described in \cite{pignatari:06}.
In particular, the $^{22}$Ne($\alpha$,n)$^{25}$Mg and 
the $^{12}$C($\alpha$,$\gamma$)$^{16}$O rates are from \cite{jaeger:01} and from
\cite{caughlan:85}, respectively.
In the initial composition, 
$\alpha-$enhancements for the light isotopes are included according to 
\cite{mishenina:00} for the oxygen and to \cite{goswami:00} and \cite{francois:04}
for the other elements. 
Heavy isotope abundances beyond iron are solar$-$scaled with the initial metallicity.\\
In Fig. \ref{m25} Top Panel the isotopic distributions between $^{57}$Fe and $^{138}$Ba 
are presented at the end of the C$-$burning shell of a 25 M$_{\odot}$ star at [Fe/H] = $-$4
for three cases:
$i$) the standard $s-$process distribution,
$ii$, $iii$) $s-$process distributions where the additional $^{22}$Ne 
is 0.5 \% and 1.0 \% in mass (about 100 and 200 times more than in the
standard case), respectively, in agreement with \cite{hirschi:08}. 
For each of these cases, in Table \ref{tab:25} the $s-$process parameters have been reported 
with the final C shell abundances Y$_{\rm i}$ for a sample of elements.
The ejected C shell mass of element i, em$_{\rm i}$,
is given by
em$_{\rm i}$ = Dm~$\times$~\={A}$_{\rm i}$~$\times$~Y$_{\rm i}$, 
where \={A}$_{\rm i}$ is the average mass number of element i and it is given 
in Table \ref{tab:25}, and Dm is the C shell mass ejected unchanged 
by the explosion in M$_{\odot}$.
For comparison, the amount of $^{56}$Fe produced
by a 25 M$_{\odot}$ star is 0.1$-$0.2 M$_{\odot}$, it is primary and 
it strongly depends on the explosion parameters \citep[e.g.,][]{rauscher:02}.\\ 
In case $i$ the isotopic distribution is peaked between Fe and Sr.
The $^{22}$Ne abundance depends on the initial CNO abundances used for a given
metallicity and on the $\alpha-$enhancements used for light elements.
As already mentioned in \S \ref{sec:intro}, for [Fe/H] $\le$ $-$2 the main neutron poisons 
are primary in the He core, and the probability that a free neutron is captured by 
iron seeds is so low that the $s-$process efficiency decreases more rapidly than the iron seeds 
with decreasing metallicity \citep[e.g.][]{raiteri:92}.
Notice that in the C shell the $s-$process efficiency is 
strongly reduced by primary neutron 
poisons like $^{20}$Ne and $^{23}$Na (produced by the $^{12}$C + $^{12}$C reaction).
At [Fe/H] = $-$4, in cases $ii$ and $iii$  
iron seeds 
are more consumed with respect to standard $s-$process calculations.
The $s-$process efficiency between Sr and Ba
is increased by orders of magnitude, producing a flat isotopic pattern in this mass region.
In particular, case $ii$ is more efficient in producing isotopes in the 
Pd region and less efficient in producing heavier elements with respect 
to case $iii$. 
On the other hand, case $ii$ and case $iii$ provide similar results for Sr, Y and Zr
(within a factor 1.3), at the 
bottle$-$neck of the neutron capture flow.
This implies that the Sr/Ba ratio decreases with increasing $^{22}$Ne:
according to Table \ref{tab:25},
in case $ii$ the Sr/Ba = 9.1 and in case $iii$ Sr/Ba = 3.8. 
The addition of more primary $^{22}$Ne does not imply a 
higher production of Sr, Y and Zr. This happens for two main reasons.
The first reason is related to the light isotope abundances.
Indeed, at [Fe/H] = $-$4 most of the neutrons produced by $^{22}$Ne are captured 
by the strongest neutron poisons, $^{25}$Mg and $^{22}$Ne
($^{16}$O is the strongest neutron absorber, but the neutrons are efficiently 
recycled via the $^{17}$O($\alpha$,n)$^{20}$Ne reaction).
The second reason is related to the $s-$process seeds.
In Fig. \ref{m25} Top Panel Fe seeds have been destroyed by the $s-$process, and the
resulting abundances for elements between Fe and Sr are smaller than the abundances of
elements beyond Sr.
In case $ii$ of Fig. \ref{m25} Top Panel the resulting $s-$process production
yields for elements up to Rb (included) is less than 1 \% than for elements
beyond it.
This implies that the neutron capture nucleosynthesis flow producing Sr is exhausted.\\
%
%\clearpage

\begin{figure}[!h]
\centering
\resizebox{18pc}{!}{\rotatebox{-90}
{\includegraphics{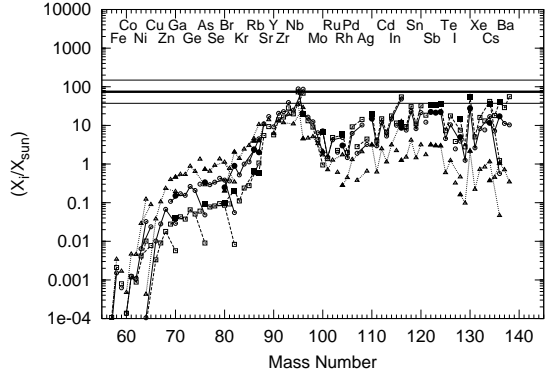}}}    
\caption{The same as case $iii$ in Fig. \ref{m25} Bottom Panel, but using different  
$^{22}$Ne($\alpha$,n)$^{25}$Mg rates: recommended (circles), recommended * 2 (squares) and
recommended / 2 (triangles).
}
\label{m25z2m5.nuc}
\end{figure}

%
%
In Fig. \ref{m25} Bottom Panel we present similar results 
at [Fe/H] = $-$3. The correspondent $s-$process parameters and
element abundances have been reported in Table \ref{tab:25}.
The new generation of models presented by \cite{hirschi:08} does not yet include models with
[Fe/H] $\sim$ $-$3.
However, since the nitrogen is observed to be primary over a large range of metallicity  
\citep[][]{spite:05}, we discuss now the possibility 
that the additional $^{22}$Ne
is produced with similar efficiency in the He core at [Fe/H] = $-$3.
In Fig. \ref{m25} Bottom Panel the isotopic distribution without additional $^{22}$Ne 
(case $i$) is similar to the isotopic distribution at solar metallicity
\citep[e.g.][]{rauscher:02,pignatari:06}. In particular, a peak of production 
is obtained in a large mass region between Cu and Sr and with a rapid fall beyond Y,
where the standard weak $s$ process is not efficient.
On the other hand, the production factors in case $i$ are about three orders of magnitude
lower than in the case at solar metallicity, confirming the secondary nature
of the weak s$-$process in massive stars.
In case $ii$ and case $iii$ the primary $^{22}$Ne 
is 0.2 \% and 1 \% in mass 
(about 10 and 50 times more than in the standard case with no rotation), respectively.
In this case we consider a larger range of $^{22}$Ne with respect to 
[Fe/H] = $-$4, to test the sensitivity of $s-$process calculations to different amounts
of the neutron source. 
Like at [Fe/H] = $-$4, the mass region Fe$-$Zn is again substantially depleted in cases $ii$ 
and $iii$ with respect to case $i$.
On the other hand, quite similar results are observed between Zn and Kr.
Finally, abundances beyond Kr are strongly enhanced in the calculations including 
primary $^{22}$Ne. 
At [Fe/H] = $-$3 the iron seeds are a factor ten higher than at [Fe/H] = $-$4,
and the $s-$process efficiency increases by about a factor ten in cases $ii$ and $iii$
with respect to similar cases in Fig. \ref{m25} Top Panel. 
As recalled above, since the neutron source $^{22}$Ne and the main 
neutron poisons $^{25}$Mg and $^{22}$Ne
are primary, the $s-$process is secondary
scaling with the iron seed abundance.
Without considering smaller local differences, the 
isotopic patterns at [Fe/H] = $-$3,$-$4 are quite similar.
Notice that changing the primary $^{22}$Ne amount by a factor five 
affects the Sr, Y and Zr production less than a factor two.
Concerning heavier isotopes, more primary $^{22}$Ne implies a more efficient
production in the Ba region, more $^{22}$Ne left in the He core ashes and as a consequence
higher neutron densities in the C
shell (e.g., $^{96}$Zr overproduction in case $iii$ with respect to case $ii$).
In particular, in case $ii$ Sr/Ba = 107.6, which is strongly larger than 
in case $iii$ with Sr/Ba = 4.6.\\
Considering metallicities higher than [Fe/H] $\sim$ $-$3 and closer to solar metallicity, 
rotational velocity and the related mixing efficiency in stars should
decrease, assuming that stars of different metallicity begin their evolution 
with similar angular momentum \citep[e.g.][]{meynet:06, hirschi:07}.
Therefore, the production of primary $^{14}$N and of the additional $^{22}$Ne
in the He core becomes negligible at solar metallicity.
This implies that the new s$-$process component discussed in this paper is  
activated at metallicities typical of the Galactic halo, but not for 
metallicities of the Galactic disk. \\
An interesting point to discuss is how much the nuclear rates affect our results.
It is well known that the $^{22}$Ne($\alpha$,n)$^{25}$Mg reaction rate and the
relative ($\alpha$,n)/($\alpha$,$\gamma$) ratio affect the $s-$process nucleosynthesis.
In Fig. \ref{m25z2m5.nuc} we present case $iii$ of Fig. \ref{m25} Bottom Panel where the 
$^{22}$Ne($\alpha$,n)$^{25}$Mg recommended rate \citep[]{jaeger:01}
has been multiplied and divided by a factor two.
Elements lighter than Sr are more produced with a lower 
$^{22}$Ne($\alpha$,n)$^{25}$Mg rate,
since less neutrons are available to feed heavier elements through following neutron captures.
The opposite effect is observed with higher $^{22}$Ne($\alpha$,n)$^{25}$Mg.
Higher ($\alpha$,n) rate and higher ($\alpha$,n)/($\alpha$,$\gamma$) ratio do not strongly affect
our calculations between Sr and Ba.
The reason is that in this case more neutrons are produced but also more $^{25}$Mg, which is the
strongest neutron poison in these conditions.
Furthermore, iron seed are destroyed and cannot feed the nucleosynthesis of Sr, Y and Zr.
For instance, in case $iii$ of Fig. \ref{m25} Bottom Panel the amount of material between Fe and
Rb is less than 1 \% of the material beyond Sr. 
On the other hand, the $s-$process distribution beyond Sr shows a strong propagation effect due to
a lower ($\alpha$,n) rate. In this case, less neutrons and $^{25}$Mg are produced but more 
$^{22}$Ne is left, which is also a neutron poison, causing a lower $s-$process efficiency.
Assuming a factor two of uncertainty for the $^{22}$Ne($\alpha$,n)$^{25}$Mg, 
the Sr/Ba ratio changes by a factor 221 (0.5 and 110.4 for the 
($\alpha$,n)*2 case and for the ($\alpha$,n)/2 case, respectively)!
Out of the range of Fig. \ref{m25} Bottom Panel, the distribution beyond Ba is also 
significantly affected.
For example, the ratio Sr/Pb ranges between 1.9 and 8059.7 for the ($\alpha$,n)*2 case
and the ($\alpha$,n)/2 case, respectively.
This large effect is due to the amount of $^{22}$Ne consumed in the He core: 
Sr, Y and Zr are no more produced, but heavier elements 
in Ba and Pb peaks are fed at the expense of lighter $s-$process isotopes.
Related to this last point is also the nuclear uncertainty of 
$^{12}$C($\alpha$,$\gamma$)$^{16}$O, which is in competition with the 
$^{22}$Ne($\alpha$,n)$^{25}$Mg capturing helium before its exhaustion in the He core.
For instance, a lower $^{12}$C($\alpha$,$\gamma$)$^{16}$O rate increases the amount of 
$^{22}$Ne burnt in the He core, and it implies a more efficient $s$ nucleosynthesis at both
the Ba and the Pb peaks.
Other nuclear uncertainties (e.g. the $^{16}$O(n,$\gamma$)$^{17}$O cross section) 
show smaller but non$-$negligible effects on the calculations.

\section{Discussion and final remarks}
\label{sec:concl}

We presented a first study of the effect of rotational mixing 
in massive stars on the $s$ process at [Fe/H] = $-$3,$-$4.
A grid of models for a large spread of masses and metallicities 
is required for a more detailed analysis, including an evaluation of 
the impact of stellar model physics uncertainties.
Notice also that other processes not considered in the present models 
may affect the mixing efficiency and the nucleosynthesis 
in massive stars, e.g., binary interaction and magnetic field 
\citep[e.g.][]{langer:08}.\\
In the calculations presented at [Fe/H] = $-$3,$-$4, the additional $^{22}$Ne 
due to rotation is assumed to be primary in the He core and in 
the following C shell,
and it causes a peak of production in the region between Sr and Ba.
In this mass region the $s-$process efficiency is orders of magnitude higher than
in the case with no rotation.
The main neutron poisons are $^{25}$Mg and $^{22}$Ne. 
Iron seeds are strongly depleted feeding heavier isotopes, 
and once they are destroyed the production flux to Sr$-$Y$-$Zr is substantially exhausted.
In these conditions, the Sr$-$Y$-$Zr $s$ nucleosynthesis is not more efficient 
increasing the $^{22}$Ne burnt in the He core,
but Sr$-$Y$-$Zr elements can feed the production of heavier elements.
The $s-$process efficiency at the Ba peak (and at the Pb peak) 
is affected by the amount of primary $^{22}$Ne, by the fraction of $^{22}$Ne 
burnt in the He core and by nuclear uncertainties, in particular by 
the $^{22}$Ne($\alpha$,n)$^{25}$Mg and the $^{12}$C($\alpha$,$\gamma$)$^{16}$O.
Indeed, the Sr/Ba ratio changes by two orders of magnitude considering a factor 
of two of uncertainty for the $^{22}$Ne($\alpha$,n)$^{25}$Mg.
The $s$ process in fast rotating massive stars could provide a new 
scenario to explain the Sr enrichment
coupled with high Sr/Ba observed in many stars \citep[][]{truran:02, aoki:05}
at very low metallicity,
including also the spread in the observed Sr/Ba ratio.
On the other hand, the $s-$process component activated by the
additional $^{22}$Ne due to rotation
is a secondary$-$like process for the Sr peak elements 
in the range of metallicity considered,
whereas the Sr enrichment is still observed in stars like
HE 1327$-$2326 ([Fe/H] = $-$5.45).
This last point could be a strong constraint to rule out this scenario.
Unfortunately, elements between Zr and Ba have been observed in few metal poor stars
to constrain the signature of the proposed $s-$process mechanism.
A noticeable exception is the star HD 122563 \citep[e.g.,][and references therein]{montes:07},
where one$-$two lines have been detected and measured for elements in this mass region.
More observations are required to draw firm conclusions.
New models at higher metallicity are necessary to study the chemical evolution of the
Galactic disk.
Models presented here could have important consequences for the chemical evolution
of elements between Sr and Ba in the Galactic halo.

%Acknowledgements:
\acknowledgments M.P. is supported by a Marie Curie Int. 
Reintegr. Grant MIRG-CT-2006-046520 within the European FP6, 
and by NSF grants PHY 02-16783 (JINA).
R.G. thanks the Italian MIUR-PRIN06 Project 2006022731\_005.  
G.M. thanks Andr\'e Maeder for enlightening discussions
on nucleosynthesis in rotating massive stars.

\clearpage
\thispagestyle{empty}
{\rotate
\begin{table*}[tp!]
\centering
%{\footnotesize
\caption{For the cases in Fig. \ref{m25} the neutron exposure $\tau$, 
the peak central neutron density in the He core and the peak neutron density 
in the C shell are reported. The initial $^{22}$Ne mass fraction available in the He core
and the final C shell number abundances Y$_{\rm i}$ for a sample of elements 
are reported, and in the first column the average mass number of element i 
is given (\={A}$_{\rm i}$).}
\label{tab:25}
\begin{tabular}{lcccccc} 
% .4\textwidth
\hline
M = 25 M$_{\odot}$  & [Fe/H] = $-$3  & [Fe/H] = $-$3 & [Fe/H] = $-$3 
& [Fe/H] = $-$4 & [Fe/H] = $-$4 & [Fe/H] = $-$4 \\
$X$($^{22}$Ne)$_{\rm ini}$ 
& 2.12~$\times$~10$^{-4}$  & 2.0~$\times$~10$^{-3}$ & 1.0~$\times$~10$^{-2}$ 
& 5.21~$\times$~10$^{-5}$  & 5.0~$\times$~10$^{-3}$ & 1.0~$\times$~10$^{-2}$ \\
\hline
$\tau$ (mb$^{-1}$)   & 0.255 & 0.659 & 0.923 &  0.118  & 0.841 & 0.942   \\
Heco(n$_{\rm n}$)$_{\rm max}$ (cm$^{-3}$)   
& 2.808~$\times$~10$^{7}$ & 6.073~$\times$~10$^{7}$ & 7.912~$\times$~10$^{7}$ 
& 1.475~$\times$~10$^{7}$ & 7.317~$\times$~10$^{7}$ & 7.976~$\times$~10$^{7}$ \\
Csh(n$_{\rm n}$)$_{\rm max}$ (cm$^{-3}$)   
& 7.278~$\times$~10$^{11}$ & 1.518~$\times$~10$^{12}$ & 3.691~$\times$~10$^{12}$ 
& 7.021~$\times$~10$^{11}$ & 2.969~$\times$~10$^{12}$ & 4.362~$\times$~10$^{12}$ \\
\hline
Sr (87.7)  & 9.304~$\times$~10$^{-11}$ & 7.440~$\times$~10$^{-9}$ & 5.595~$\times$~10$^{-9}$ 
& 5.823~$\times$~10$^{-13}$ & 6.589~$\times$~10$^{-10}$ & 5.294~$\times$~10$^{-10}$ \\
Y (89.0)   & 6.964~$\times$~10$^{-12}$ & 1.746~$\times$~10$^{-9}$ & 1.992~$\times$~10$^{-9}$ 
& 6.450~$\times$~10$^{-14}$ & 2.130~$\times$~10$^{-10}$ & 1.935~$\times$~10$^{-10}$ \\
Zr (91.3)  & 5.754~$\times$~10$^{-12}$ & 2.806~$\times$~10$^{-9}$ & 5.094~$\times$~10$^{-9}$ 
& 6.623~$\times$~10$^{-14}$ & 4.745~$\times$~10$^{-10}$ & 5.115~$\times$~10$^{-10}$ \\
Mo (96.0)  & 3.751~$\times$~10$^{-13}$ & 3.656~$\times$~10$^{-10}$ & 6.885~$\times$~10$^{-10}$ 
& 6.663~$\times$~10$^{-15}$ & 1.055~$\times$~10$^{-10}$ & 1.526~$\times$~10$^{-10}$ \\
\hline
Ba (137.4) & 5.222~$\times$~10$^{-13}$ & 6.917~$\times$~10$^{-11}$ & 1.222~$\times$~10$^{-9}$ 
& 3.967~$\times$~10$^{-14}$ & 7.270~$\times$~10$^{-11}$ & 1.378~$\times$~10$^{-10}$ \\
\hline
\hline
\end{tabular}
%}
\end{table*}}

\clearpage

\end{document}